\documentstyle[epsfig]{article}
\begin{document}
% \begin{frontmatter}
\title{Evidence for Different Freeze-Out Radii of High- and 
       Low-Energy Pions Emitted in Au+Au Collisions at
       1~A$\cdot$GeV}
%----------------------------------------------------------------
%\title{Evidence for Different Freeze-Out Radii of High- and 
%       Low-Energy Pions Emitted in Au+Au Collisions at
%       1~A$\cdot$GeV}
%----------------------------------------------------------------
%---- Author's address                                       ----
%----------------------------------------------------------------
% Andreas Wagner\\
% c/o Gesellschaft f\"ur Schwerionenforschung\\
% Postfach 11 05 52\\
% D-64220 Darmstadt
%
% FAX:   +49-6159-71-2809\\
% Phone: +49-6159-71-2448\\
%----------------------------------------------------------------
%---- Author and coauthors                                   ----
%----------------------------------------------------------------
\author{(A.~Wagner, C.~M\"untz, H.~Oeschler, C.~Sturm)\footnote{
          Technische Universit\"at Darmstadt, 
          D-64289~Darmstadt, Germany}\\
        (R.~Barth, M.~Cie\'slak, M.~D\c{e}bowski, 
         E.~Grosse, \\ P.~Koczo{\'n}, M.~Mang, D.~Mi\'skowiec,
         R.~Schicker, \\ E.~Schwab, P.~Senger)\footnote{
          Gesellschaft f\"ur Schwerionenforschung, 
          D-64291~Darmstadt, Germany}\\
        (P.~Beckerle, D.~Brill, Y.~Shin, H.~Str\"obele)\footnote{
          Johann Wolfgang Goethe-Universit\"at, 
          D-60054~Frankfurt, Germany}\\
        (W.~Walu\'s)\footnote{
          Uniwersytet Jagiello\'{n}ski, 
          PL-30-059~Krak\'ow, Poland}\\
        (B.~Kohlmeyer, F.~P\"uhlhofer, J.~Speer, 
         K.~V\"olkel)\footnote{
          Phillips-Universit\"at Marburg,
          D-35037~Marburg/Lahn, Germany}\\}
\maketitle  
\begin{abstract}
Double differential production cross sections of $\pi^-$ and
$\pi^+$ mesons and the number of participating protons have been measured
in central Au+Au collisions at 1~A$\cdot$GeV.
At low pion energies the $\pi^-$ yield is strongly enhanced
over the $\pi^+$ yield.
The energy dependence of the $\pi^-/\pi^+$ ratio is 
assigned to the Coulomb interaction
of the charged pions with the protons in the reaction zone.
The deduced Coulomb potential increases with increasing pion c.m.~energy.
This behavior indicates different freeze-out radii for different 
pion energies in the c.m.~frame.
\end{abstract}
% \end{frontmatter}
%----------------------------------------------------------------
%---- PACS number 25.75.-q, 25.75.Dw                         ----
%----------------------------------------------------------------
%---- Introduction                                           ----
%----------------------------------------------------------------
Relativistic heavy ion collisions are a unique tool to study 
nuclear matter at high densities and temperatures~\cite{Stock:86}.
At bombarding energies around 1~A$\cdot$GeV, baryonic densities
of 2--3 times normal nuclear matter density 
($\rho_0 = 0.17~\mbox{fm}^{-3}$)
are expected to be reached during a time span of 10-15~fm/c 
(see e.g.:~\cite{Aichelin:91}).
Thereafter, the high pressure built up in the reaction zone 
drives the system apart and particles freeze out according to their
mean free path.
A major experimental challenge is to determine the baryonic 
density in the different stages of the reaction.

This determination requires to measure both the number of baryons 
inside the reaction zone and its time-dependent volume.
The number of participating nucleons is derived from the projectile 
spectator charges~\cite{Brill:96},
whereas the measurement of their spatial extent is more complex.
Two-particle correlations ($ \pi\pi, KK, pp$) 
can be used to extract information on the size
and lifetime of the source at freeze-out~\cite{Boal:90,Hote:92}. 

As an alternative approach, presented in this Letter,
the Coulomb potential of the nuclear fireball is extracted by studying 
its interaction with oppositely charged pions.
The strength of the Coulomb field depends on the number of
participating protons and their geometrical configuration
(resp. the freeze-out radius) at the time of pion emission.
At a beam energy of 1~A$\cdot$GeV pion production
mainly proceeds via the excitation and decay of the 
$ \Delta_{33}(1232)$-resonance.
In this case the overall $\pi^-/\pi^+$-ratio is determined by the 
$N/Z$-ratio of the colliding nuclei via isospin conservation.
The variation of the $\pi^-/\pi^+$-ratio as a function of the 
pion energy is attributed to the Coulomb field of 
the charge in the reaction volume which de- or accelerates the 
pions according to their charge.
Previously, the influence of Coulomb effects on the emitted pions 
has been studied for different reaction systems and 
energies~\cite{Miller:87,Harris:90} without 
deducing the fireball's Coulomb field. 
For example, in Ne+NaF collisions at 0.4~A$\cdot$GeV
a strong peak in the $\pi^-/\pi^+$-ratio has been found at laboratory 
angles close to zero degrees which was explained by Coulomb interaction 
of the charged pions with the projectile spectator~\cite{Benenson:79}.

%----------------------------------------------------------------
%---- Experimental setup                                     ----
%----------------------------------------------------------------
In this Letter we report on the measurement of 
double differential production cross sections of positively
and negatively charged pions 
emitted in central $\rm^{197}Au$+$\rm^{197}Au$ collisions at
1~A$\cdot$GeV incident kinetic energy impinging on a
target of 1.93~g/cm$^{-2}$.
The experiment was performed with the Kaon spectrometer~\cite{Senger:93}
at the heavy-ion synchrotron SIS at GSI. %(Darmstadt, Germany).
The spectrometer covers a momentum-dependent solid angle of
$\Delta\Omega$~=~(15-35)~msr with a momentum resolution of
$\delta p/p \simeq 0.01$ over the full momentum range.
The momentum resolution is dominated by multiple scattering.
The momentum acceptance is $p_{max}/p_{min} \approx 2$ 
for a given magnetic field setting.
The measured laboratory momenta vary between 
0.156~GeV/c and 1.5~GeV/c.
The spectra were obtained at laboratory angles of
($44\pm 4$)~degrees.
This angular range corresponds to a range of normalized rapidity 
of $0.43 \leq y/y_{beam} \leq 0.74$ with transverse momenta
between 0.12~GeV/c and 1.2~GeV/c.
The particle identification is performed by the reconstruction of the
trajectory and by the determination of the particle velocity
with two time-of-flight arrays.
%------------NEW-----------
The collision centrality is determined by means of the hit
multiplicity of charged particles in a 84-fold segmented scintillation 
detector close to the target covering polar angles between 12 and 48 degrees.
The corresponding number of participating protons is derived from the
summed charge of projectile spectator fragments measured at forward
angles by a 380-fold segmented hodoscope. The method is
discussed in ref.~\cite{Brill:96}.
In order to minimize the influence of spectator matter 
(nucleons not contributing to the reaction zone)
we have selected collisions with the highest 
charged particle multiplicity and 
pion emission around mid-rapidity.
These central collisions which were selected represent 
$(14\pm4)~\%$ of the total reaction cross section.
The corresponding number of participating protons is determined
to be $<Z_{part}> = 110\pm8$.
%--------end-----NEW

\begin{figure}
\epsfig{file=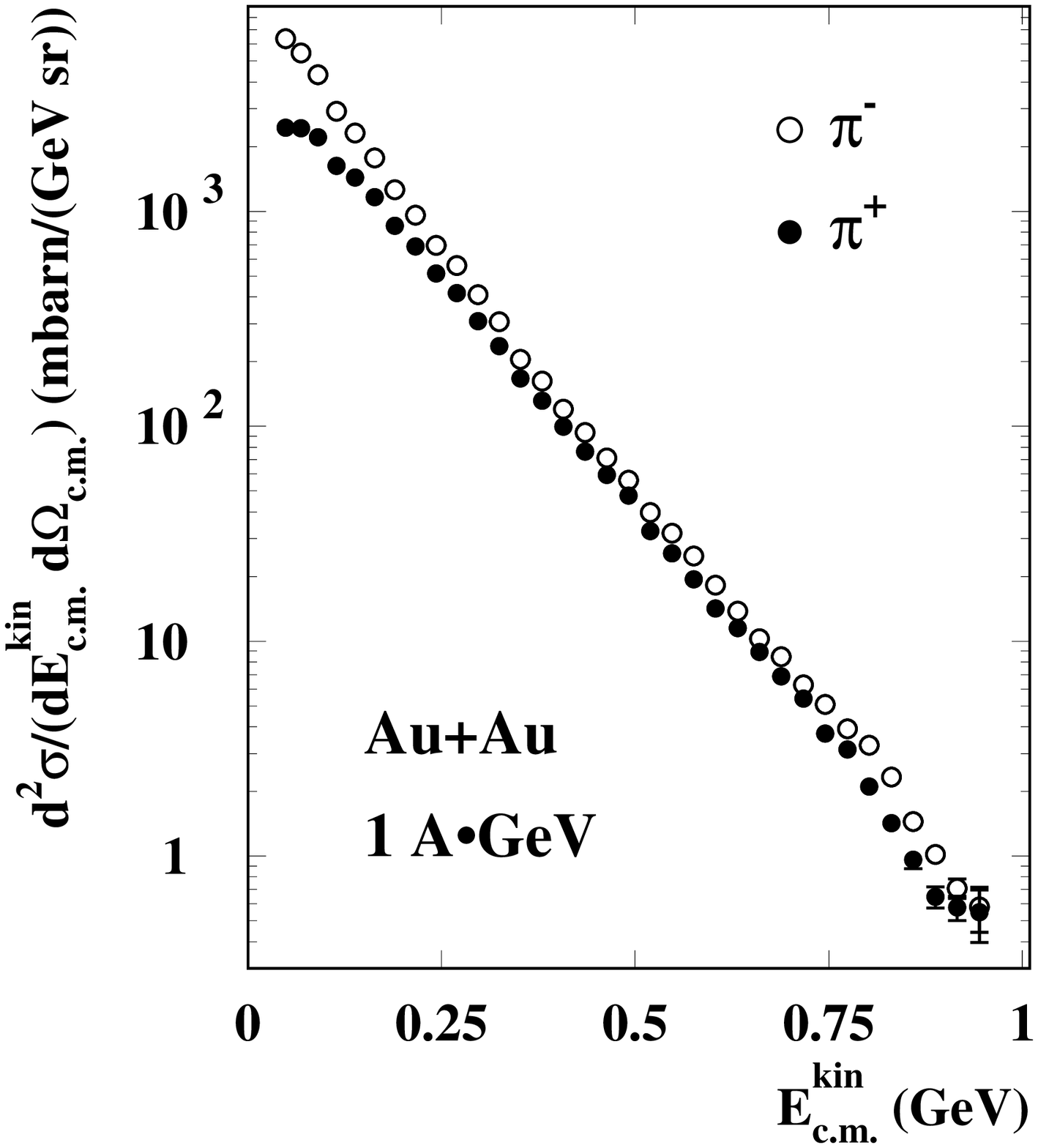,width=13cm}
\caption{Production cross section of negatively and positively 
         charged pions from central collisions 
         of the reaction system $ ^{197}$Au+$^{197}$Au
         at an incident beam energy of 1~A$\cdot$GeV and 
         at an emission angle of $ \theta_{LAB} = (44\pm4)$ degrees.}
\label{plo_r_tsys}
\end{figure}

Figure~\ref{plo_r_tsys} shows the $\pi^-$ and $\pi^+$ cross section 
$d^2\sigma/(dE_{c.m.}^{kin} d\Omega_{c.m.})$
as a function of the kinetic energy $E_{c.m.}^{kin}$ in the 
center-of-momentum frame for central Au+Au reactions.
The $\pi^-$ yield clearly exceeds the $\pi^+$ yield at low kinetic 
energies 
whereas at higher energies these two yields are close to each other.
Similar results are obtained by the FOPI collaboration, yet over a smaller 
momentum byte~\cite{Pelte:97}. 

In~order to determine the energy-integrated
$\pi^-/\pi^+$-ratio
$R_{exp} = (d\sigma(\pi^-)/d\Omega)/(d\sigma(\pi^+)/d\Omega)$
we parameterized the pion spectra by the sum of two
Maxwell-Boltzmann distributions and 
extrapolated the fit into the non-measured energy range 
(for details see ref.~\cite{Muentz:97}).

The overall $\pi^-/\pi^+$-ratio of 
$ R_{exp} = 1.94\pm0.1 $  
agrees well with the ratios derived from the isospin decomposition of 
individual nucleon-nucleon collisions at the given beam energy:
Using the parameterization of ref.~\cite{VerWest:82}
one obtains a value of $ R_{iso} = 1.90$.
The corresponding value from the isobar model which assumes a
pion production exclusively via a isospin 3/2 resonance is 
$ R_{\Delta} = 1.95$.
The agreement between experimental and theoretical values 
indicates that the $ \pi^-/\pi^+$-ratio 
reflects the N/Z asymmetry of the colliding system.

\begin{figure}
\epsfig{file=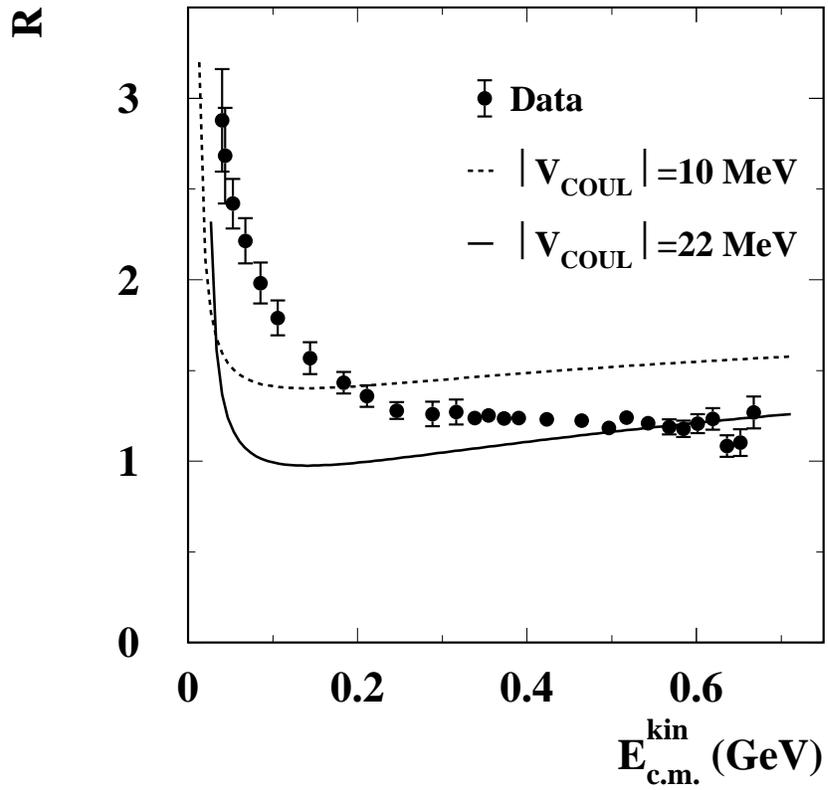,width=13cm}
\caption{$\pi^-$/$\pi^+$ ratio for the reaction system 
         $^{197}$Au+$^{197}$Au 
         as a function of the pion kinetic energy
         for central collisions.
         The solid (dotted) curve shows the calculated ratio for a 
         fixed Coulomb energy of 22~MeV (10~MeV).
         The error bars reflect statistical errors only; 
         a systematic error of $\pm$5\% has to be added.}
\label{plo_r_rsys}
\end{figure}

%----------------------------------------------------------------
%---- Energy dependent particle ratios                       ----
%----------------------------------------------------------------
Figure~\ref{plo_r_rsys} shows the $ \pi^-/\pi^+$-ratio
for Au+Au as a function of the pion kinetic energy in the
c.m. system.
The $ \pi^-/\pi^+$-ratio decreases from 2.8 at low pion energies to
a nearly constant value of 1.1 for pion energies above 0.4~GeV.

For the following analysis we assume that the $ \pi^-/\pi^+$-ratio
is independent of the pion energy if Coulomb effects are 
disregarded.
This assumption is supported by transport model calculations
(IQMD~\cite{Bass:95}, BUU~\cite{Teis:97}) which find - within the
error bars - a constant $ \pi^-/\pi^+$-ratio of
$ \approx 1.9$ for all pion energies if the Coulomb interaction is
switched off.
The BUU calculation % model 
also includes baryonic resonances heavier than the 
$\Delta_{33}(1232)$ resonance.

%----------------------------------------------------------------
%---- Derivation of the Coulomb potential                    ----
%----------------------------------------------------------------
In order to extract the strength of the Coulomb force
we use a static approximation for the Coulomb field.
This assumption is justified as we consider only pion kinetic 
energies above 0.14~GeV. Their velocities ($\beta > 0.87$) are significantly
larger than the expansion velocity of the nuclear 
fireball ($\beta \approx 0.4$).
(Recently, 
the influence of an expanding source has been studied
for charged kaons showing a negligible effect at higher kinetic 
energies~\cite{Ayala:97}.)
According to ref.~\cite{Gyulassy:81}, the Coulomb force
distorts the pion spectra by modifying both the kinetic energies of the
particles and the available phase space.
One derives the following relations for the measured cross sections
at a given momentum $ p$
\begin{equation}
 \sigma(\vec{p}) \equiv
 \frac{\displaystyle d^3\sigma}{\displaystyle dp^3} =
 \sigma_0(\vec{p_0}(\vec{p})) \cdot
 \left|\frac{\displaystyle \partial^3 p_0}{\displaystyle\partial^3 p} \right|
 \label{equ:sigma}
\end{equation}
with the undisturbed momentum $\vec{p_0}$
and the undisturbed differential cross section
$\sigma_0(\vec{p_0}(\vec{p}))$
taken at the momentum $\vec{p_0}(\vec{p})$.

The Jacobian in eq.~(\ref{equ:sigma}) (Coulomb phase-space factor) is
evaluated assuming spherical symmetry with the relativistic 
energy-momentum relation and by using the identity 
$ p \partial p = E \partial E $:
\begin{equation}
 \left|\frac{\displaystyle\partial^3 p_0}
            {\displaystyle\partial^3 p} \right| =
 \left|\frac{\displaystyle p_0 E_0 \partial E_0}
            {\displaystyle p E \partial E} \right| =
       \frac{\displaystyle p_0 E_0}
            {\displaystyle p E}.
\label{equ:jacobian}
\end{equation}
As the pion  spectra are not described by a single
Maxwell-Boltzmann distribution, the 
undisturbed differential cross section is parameterized by
$ d^3\sigma_0 / dp^3_0 \propto \exp(- \beta(E_0) \cdot E_0)$ 
with $\beta^{-1} = 32~\mbox{~MeV} + 7.4 \cdot 10^{-2} \cdot E_0$.
This function has been obtained as the average of the functions $\beta(E_0)$
obtained for the measured $\pi^-$ and $\pi^+$ spectra.
For inclusive reactions the function $\beta$ determined in this way
describes well the curvature 
of neutral pion spectra measured in the same reaction~\cite{Schwalb:94}.
     
With (c=1)
\begin{equation}
          E_\pm  =  E_0 \pm V_{Coul}; \qquad p_\pm = \sqrt{E_\pm^2 - m^2}
\end{equation}
one writes for the spectra of charged pions
\begin{equation}
  \sigma_\pm(\vec{p_\pm}) \equiv 
  f_\pm \cdot \sigma_0(\vec{p_0}(\vec{p_\pm})) \cdot 
  \frac{\displaystyle p_0 E_0}
       {\displaystyle p_\pm E_\pm}.\\
\end{equation}
The plus (minus) sign refers to $ \pi^+ (\pi^-)$,
and $f_\pm$ reflects the weights due to the isospin coefficients. 
With $R^{tot}_{iso} \equiv f_-/f_+ $ 
the $\pi^-/\pi^+$ ratio can be written as
\begin{equation}
  R = \frac{\displaystyle \sigma_-(\vec{p_{-}})}
           {\displaystyle \sigma_+(\vec{p_{+}})} =
  R^{tot}_{iso} \cdot 
         \frac{\sigma_0(\vec{p_0}(\vec{p_{-}}))}
              {\sigma_0(\vec{p_0}(\vec{p_{+}}))}
         \cdot 
         \frac{\displaystyle p_{+} E_{+}}
              {\displaystyle p_{-} E_{-}}.
 \label{equ:vcoul}
\end{equation}
It is worth noticing that for $ p \rightarrow \infty$ the value of
$ R$ does not necessarily approach unity. %one.

In a first attempt we determine the Coulomb energy $ V_{Coul}$
using eq.~(\ref{equ:vcoul}).
%----------------------------------------------------------------
%---- Varying the Coulomb potential                          ----
%----------------------------------------------------------------
However, with the assumption of a constant Coulomb potential $ V_{Coul}$ 
one cannot describe the $\pi^-$/$\pi^+$ ratio in the entire energy range.
According to eq.~(\ref{equ:vcoul}), a $\pi^-$/$\pi^+$ ratio of around 1.1 as
measured for pion kinetic energies around 0.6~GeV requires a
Coulomb potential of 22~MeV.
This value, however, is at variance with the $\pi^-$/$\pi^+$ ratio 
at lower pion energies (see the solid line in fig.~\ref{plo_r_rsys}).
Here, the observed ratio clearly exceeds the solid line suggesting 
a smaller value for the Coulomb potential at lower pion energies.
Other attempts to describe the measured $\pi^-$/$\pi^+$ ratio
with a constant $ V_{Coul}$ failed as well~\cite{Li:94,Gorenstein:97}.

One can estimate the freeze-out radius of high-energy 
pions ($E^{kin}_{c.m.}$~=~0.6~GeV) from the determined
Coulomb potential of 22~MeV.
We assume that the pions are emitted from the surface of a charged sphere,
containing all participating protons.
The assumption of surface emission is justified by the large pion
re-absorption probability inside the reaction volume.
The Coulomb potential is then given by
$V_{Coul} = Z_{part}\cdot e^2\cdot <r^{-1}>$.
%-------NEW--------
With $<Z_{part}> = 110\pm8$  and
%--end----NEW---------
with $V_{Coul}$~=~22~MeV 
one obtains an effective radius of
\begin{equation}
 r_{eff} \equiv <r^{-1}>^{-1} = (7.2\pm1.1)~\mbox{fm}. \nonumber
\end{equation}
According to
$ \rho_{eff} = 3 A_{part} / (4 \pi r_{eff}^3)$
with the number of participating nucleons $ A_{part}$
this value corresponds to an effective density of
$\rho_{eff} = (1.1_{~-0.3}^{~+0.2})\cdot\rho_0.$

\begin{figure}[ht]
\epsfig{file=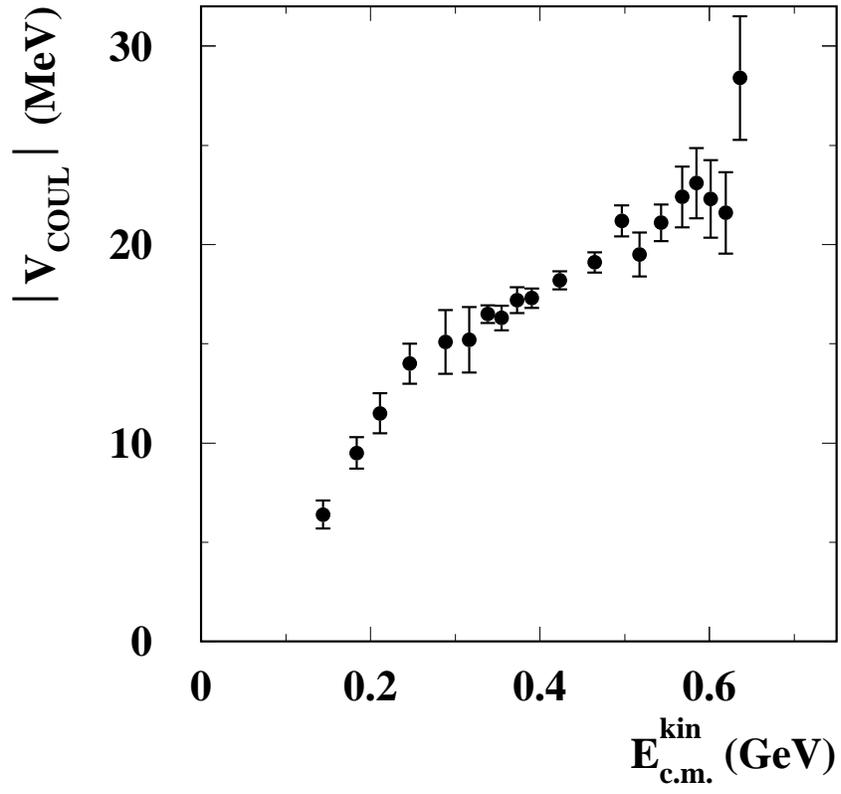,width=13cm}
\caption{Variation of $ \vert V_{Coul} \vert$ with the pion kinetic energy 
         as deduced from the measured $ \pi^-$/$\pi^+$ ratio for central 
         $ ^{197}$Au+$^{197}$Au collisions.}
\label{plo_r_vc45}
\end{figure}

In the following step we determine the Coulomb potential for 
lower pion energies.
The disagreement between the solid line 
(calculated with a constant Coulomb potential ) 
and the data in fig.~\ref{plo_r_rsys}
is a signature for a different Coulomb field acting on low-energy pions.
In order to describe the measured ratio $R$ over the entire energy range 
we vary in eq.~(\ref{equ:vcoul}) the Coulomb potential
$ V_{Coul}$ as a function of the pion kinetic energy.
The phase space factor is modified accordingly.
The resulting dependence of $ V_{Coul}$ on $ E_{c.m.}^{kin}$ is shown
in fig.~\ref{plo_r_vc45}:
$V_{Coul}$ changes from about %values around
22~MeV for high-energy pions to less than 10~MeV for
pion energies below 0.2~GeV.
At even lower pion kinetic energies pions 
have de~Broglie wavelengths comparable to the size of the
emitting source; this problem requires treatment 
in terms of quantum dynamics which is beyond the scope of this letter.

A value of 10~MeV for pion energies below 0.2~GeV corresponds to an effective 
freeze-out radius of $ r_{eff} = (16\pm2)$~fm and an effective density 
of $\rho_{eff} = (0.1 \pm 0.03)\cdot\rho_0$.
The reduction of $ V_{Coul}$ therefore indicates a more dilute charge
distribution at freeze-out for low-energy pions and consequently 
larger effective freeze-out radii.

A dilute nucleon distribution is expected for the late stage of the 
reaction whereas higher densities are reached in an earlier stage.
This interpretation is
in agreement with microscopic transport models which
predict that high-energy pions (transverse momenta $p_T \geq$~0.5~GeV/c)
are emitted during an early stage of the
collision ($t \leq 15$~fm/c) where the central baryon density is 
above $2 \cdot \rho_0$ 
while low-energy pions are emitted later during expansion~\cite{Li:91,Bass:94}.
This scenario is also supported by the experimental observation, that the 
multiplicity of high-energy pions ($E^{kin}_{c.m.} \geq$~0.45~GeV)
increases more than linearly with the number of participating 
nucleons~\cite{Muentz:95}.
This behavior is a signature of pion production via multiple baryon-baryon 
collisions which occur more frequently at higher densities.
For the system Ar+KCl at an incident energy of 1.5~A$\cdot$GeV
a variation of the source size with the pion energy has been
found previously from studying two-pion correlations~\cite{Beavis:83}.

%----------------------------------------------------------------
%---- Experiments at higher energies and their drawbacks     ----
%----------------------------------------------------------------
Recently $\pi^-$/$\pi^+$-ratios have also been measured at much
higher bombarding energies.
In Au+Au collisions at 10.7~A$\cdot$GeV the 
$\pi^-$/$\pi^+$-ratio is found to be 1.5 for low transverse 
momenta $ p_T$ and it approaches unity at higher 
$ p_T$~\cite{Ahle:96}.
In Pb+Pb collisions at 158~A${\cdot}$GeV the
$\pi^-$/$\pi^+$-ratio 
exceeds one only 
for very small values of $ p_T$.
This weak energy-dependence is explained by Coulomb interaction 
with a rather small
co-moving charge of $ Z_{eff} = 40$~\cite{Boggild:96,Osada:96}.
Two features of the measurements at the higher incident energies differ from
our observations: at high bombarding energies the number of pions per
baryon is significantly higher and due to charge conservation
the overall $\pi^-$/$\pi^+$-ratio tends to be close to 
unity~\cite{Gorenstein:95}.
The small energy dependence reveals
a rather weak Coulomb effect (low number of charges $ Z_{eff}$)
likely due to a fast disintegration of the source.
In contrast, the large variation of $\pi^-$/$\pi^+$ from 2.8 
to about 1.1 at high pion energies as found in the Au+Au system at 
1~A$\cdot$GeV
(see fig.~\ref{plo_r_rsys}) % gives evidence for 
indicates a strong Coulomb effect
due to a slowly expanding source (compared to the pion velocity)
with a large effective charge $ Z_{eff}$. 

%----------------------------------------------------------------
%---- Summary and conclusion                                 ----
%----------------------------------------------------------------
In summary, we have measured $\pi^-$ and $\pi^+$ spectra
in central Au+Au collisions at 1~A${\cdot}$GeV
together with the number of participating protons.
For this heavy system we found a strong enhancement of the 
$\pi^-$ yield over the $\pi^+$ yield at low pion kinetic energies.
Based on the assumption that the $\pi^-/\pi^+$ ratio is given 
by Coulomb interaction and
isospin conservation, the Coulomb potential of the reaction volume
is determined.
With the additional experimental information on the number of 
participating protons,
effective source radii are derived.
They are found to be smaller
for high-energy pions than for low energy pions.
This result can be interpreted as a consequence of a pion emitting 
source which expands: high-energy pions are emitted more 
likely in the dense and hot phase of the reaction
while low-energy pions leave the system after expansion 
at a more dilute phase.

%----------------------------------------------------------------
%---- Thanks                                                 ----
%----------------------------------------------------------------
We wish to thank M.~Gyulassy (Columbia Univ., New York) 
for valuable discussions.
This work is supported by the 
German \emph{Bundesministerium f\"ur Bildung und Wissenschaft,
Forschung und Technologie} under contract 06~DA~473, by the 
Polish Committee of Scientific Research under 
contract 2P03B11109, and by the
\emph{Gesellschaft f\"ur Schwer\-ionen\-forschung} under contract DA~OEK.

%----------------------------------------------------------------
%---- Bibliography                                           ----
%----------------------------------------------------------------

%----------------------------------------------------------------
%---- Figures and figure captions                            ----
%----------------------------------------------------------------
\newpage

%----------------------------------------------------------------
%---- End of the text                                        ----
%----------------------------------------------------------------
\end{document}